# Parametric Generalized Adaptive Moment Features (PG-AMF) for Bearing Fault Diagnosis and Machine Health Monitoring

Rajeev Kumar
*Department of Mechanical Engineering*
*University of Alberta*
Edmonton, Canada
rajeev4@ualberta.ca

***Abstract*— Accurate fault diagnosis of rolling element bearings in rotating machinery is considered essential for ensuring industrial safety and enabling predictive maintenance. Conventional statistical feature-based methods rely on predefined descriptors, whose diagnostic sensitivity is constrained by fixed configurations and limited adaptability across varying fault conditions. Although deep learning approaches offer strong representational capacity, their effectiveness is often restricted by high data requirements and reduced interpretability. In this work, a parametric adaptive feature extraction framework is proposed, in which feature characteristics are learned directly from data rather than being manually specified. Multiple complementary representations are extracted from vibration signals, including absolute features capturing signal energy distribution, signed moment features reflecting waveform asymmetry, and AC-coupled moment features emphasizing dynamic fluctuations, while interactions between multiple sensor channels are modeled through a structured fusion mechanism to enhance fault representation. The proposed approach is evaluated on a benchmark gearbox bearing dataset comprising five health conditions, including normal operation and multiple fault types. Improved classification performance is observed compared to conventional methods, with consistent results under cross-validation, indicating strong generalization capability. Additionally, enhanced feature separability is demonstrated through clearer clustering patterns in low-dimensional projections. The learned representations effectively capture a wide range of signal characteristics, supporting both improved diagnostic performance and practical applicability in industrial monitoring systems.**

## I. Introduction

Rolling element bearings are among the most failure-prone components of rotating machinery in industrial environments including wind turbines, conveyor systems, and electric motors. Surveys by the IEEE Industrial Applications Society and the Japan Electrical Manufacturers Association estimate that bearing faults account for 30–40% of all rotating machinery failures [1]. Undetected bearing degradation can escalate into catastrophic mechanical failure, causing unplanned downtime, safety hazards, and significant economic loss. Early and accurate fault diagnosis is therefore a key enabler of condition-based and predictive maintenance strategies.

Vibration-based fault diagnosis has been extensively studied due to the diagnostic richness of accelerometer signals [2]. Traditional approaches rely on hand-crafted statistical time-domain descriptors: root mean square (RMS) energy captures amplitude, kurtosis detects impulsive events, and skewness quantifies signal asymmetry [3], [4]. While computationally efficient, these descriptors are defined by fixed integer or rational exponents that may not optimally discriminate all fault classes. For instance, kurtosis (fourth-order moment) is known to be sensitive to transient impulses but can be degraded by outliers and is insensitive to sub-impulsive or compound faults [5]. The fixed nature of these features represents a fundamental limitation: no single exponent set is universally optimal across all fault types, operating conditions, or machine configurations.

Deep learning methods, particularly convolutional neural networks (CNNs) applied to raw signals or time-frequency representations, have demonstrated strong classification performance [6], [7]. However, CNN-based models typically require large labeled training sets, involve many parameters, and produce features that are difficult to interpret in relation to established signal processing theory [8]. Spectrogram-based CNN approaches introduce additional preprocessing complexity and are sensitive to transform hyperparameters [9]. Hybrid architectures combining CNN, Bi-LSTM, and multi-head self-attention have been proposed to address temporal and spatial feature dependencies [10], yet the interpretability of the resulting latent features remains limited. The research gap lies in developing a feature extraction framework that is (i) adaptive to the fault structure of the data, (ii) parameter-efficient, (iii) interpretable in terms of classical signal statistics, and (iv) robust to limited labeled data.

The above gap is addressed in this work by introducing the PG-AMF framework, which replaces fixed moment exponents with learnable scalar parameters $\alpha$ optimized through gradient descent. The exponents are regularized to maintain diversity across features and compactness within classes. Cross-channel information is fused via bilinear interaction and gated residual connections, enabling the model to capture complementary directional vibration signatures without resorting to deep convolutional stacks. The primary contributions of this paper are summarized as follows:

- A PG-AMF framework is proposed in which moment exponents are parameterized as learnable scalars $\alpha$, enabling data-driven adaptation of feature sensitivity across Normal, Ball, Inner Race, Compound, and Outer Race fault classes.

- Three moment families (absolute, signed, AC-coupled) are defined per channel, providing multi-scale characterization of signal amplitude, asymmetry, and fluctuation.

- A cross-channel bilinear fusion module with gated residual connections is introduced to capture directional interaction between horizontal and vertical accelerometer channels.

- A composite training objective combining focal cross-entropy, diversity margin loss, and cosine compactness

regularization is formulated to prevent feature collapse and improve class separability.

- Systematic evaluation on the XJTU Gearbox bearing dataset demonstrates improved accuracy, Macro-F1, and FDR relative to a fixed statistical baseline, with interpretable learned exponents linking to classical moment theory.

The remainder of this paper is organized as follows. In Section II, the related work is reviewed. In Section III, the dataset and experimental setup are described. In Section IV, the proposed PG-AMF architecture is presented. In Section V, the experimental results are reported and the findings are discussed, including cross-validation analysis. Finally, in Section VI, the paper is concluded.

## II. Related Work

### A. Statistical Feature Methods

Time-domain statistical features have formed the foundation of vibration-based fault diagnosis for decades. RMS, peak-to-peak, crest factor, kurtosis, and skewness are routinely extracted from raw vibration segments [3], [4]. Prasad et al. [11] proposed a multiscale statistical moment (MSM) analysis combined with a sparse autoencoder, demonstrating improved sensitivity to incipient faults compared to fixed second- and fourth-order moments. Lei et al. [12] showed that kurtosis-based health indicators effectively partition full bearing life cycles, though at the cost of reduced discriminability between structurally similar fault types such as Inner Race and Compound faults. A key limitation of all fixed-exponent approaches is that the chosen order implicitly assumes a specific signal distribution model, which may not align with the data-generating process for compound or low-severity faults [5], [11].

### B. Deep Learning Approaches

CNN-based architectures operating on raw one-dimensional vibration signals have been widely adopted for bearing fault classification [6], [13]. Fu et al. [6] constructed a parallel CNN-LSTM network combining temporal and time-frequency features extracted via continuous wavelet transform (CWT), achieving competitive accuracy on standard benchmarks. Guo et al. [14] integrated attention mechanisms with Bi-LSTM to enhance sensitivity to fault-relevant signal segments. Siddique et al. [10] combined CWT, multi-head self-attention, Bi-LSTM, and 1D-ResNet, reporting strong generalization on the Paderborn bearing dataset. Rehman et al. [8] conducted a critical survey of bearing fault CNN methods, identifying common limitations: data-intensive training, sensitivity to domain shift, and opacity of learned representations. Transformer-based models have recently been explored [15], but their large parameter counts constrain deployment in embedded industrial systems.

### C. Higher-Order Moment and Adaptive Feature Methods

Higher-order statistics, particularly spectral kurtosis and its variants including the kurtogram and L-kurtosis, have been shown to improve fault detection under noise [5], [16]. Zheng et al. [17] applied the EMDOS-DCCNN model for variable-speed bearing diagnosis, coupling empirical mode decomposition with a dual-channel CNN. Kolmogorov-Arnold network-based classifiers have recently been explored to provide explainable fault diagnosis with learnable activation functions [18]. Despite these advances, no existing work frames moment exponents as continuously differentiable learnable parameters optimized end-to-end within a compact MLP pipeline. The proposed PG-AMF framework fills this gap by unifying adaptive exponent learning with multi-family moment extraction, cross-channel fusion, and interpretable training regularization.

## III. Dataset and Experimental Setup

### A. XJTU Gearbox Bearing Dataset

The XJTU Gearbox dataset [19], collected at Xi'an Jiaotong University, provides vibration signals from a planetary gearbox test platform (Fig. 1). The test rig consists of a 3-phase, 3-HP driving motor, a planetary gearbox, a parallel gearbox, and a magnetic brake. Two PCB 352C04 uniaxial accelerometers are mounted in the horizontal (Ch1) and vertical (Ch2) directions on the planetary gearbox casing. Signals are acquired at a sampling frequency of 20,480 Hz with the motor speed set to 1,800 r/min. Five bearing health states are considered: Normal condition, Ball fault, Inner Race fault, Compound fault (concurrent inner, outer, and ball defects), and Outer Race fault. Raw recordings contain approximately 2.3 million samples per class, yielding a total of 11.68 million samples across all five classes. The dual-channel configuration provides complementary directional vibration information and motivates the cross-channel fusion module described in Section IV.

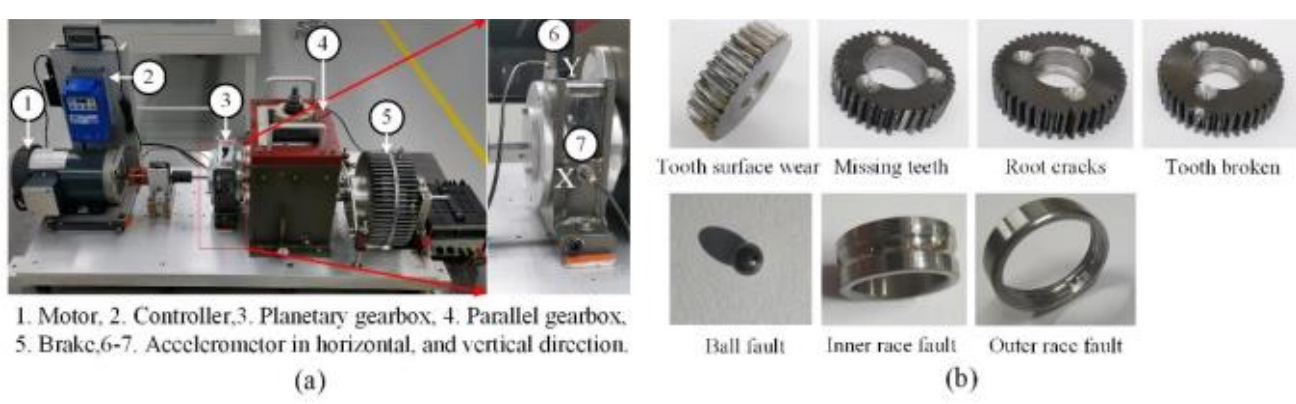

Fig. 1. XJTU Gearbox test rig (left) and health conditions of gear and bearings (right) [19].

### B. Signal Segmentation and Preprocessing

Fixed-length segments of 8,192 samples (approximately 400 ms at 20,480 Hz) are extracted with 50% overlap. Each segment is mean-subtracted and normalized to zero mean and unit variance to remove DC bias and amplitude scaling effects. A balanced set of 1,000 segments (200 per class) is drawn for all experiments. The final dataset tensor has shape (1000, 2, 8192), representing samples, channels, and segment length. A stratified split of 65/20/15 yields 650 training, 200 validation, and 150 test segments.

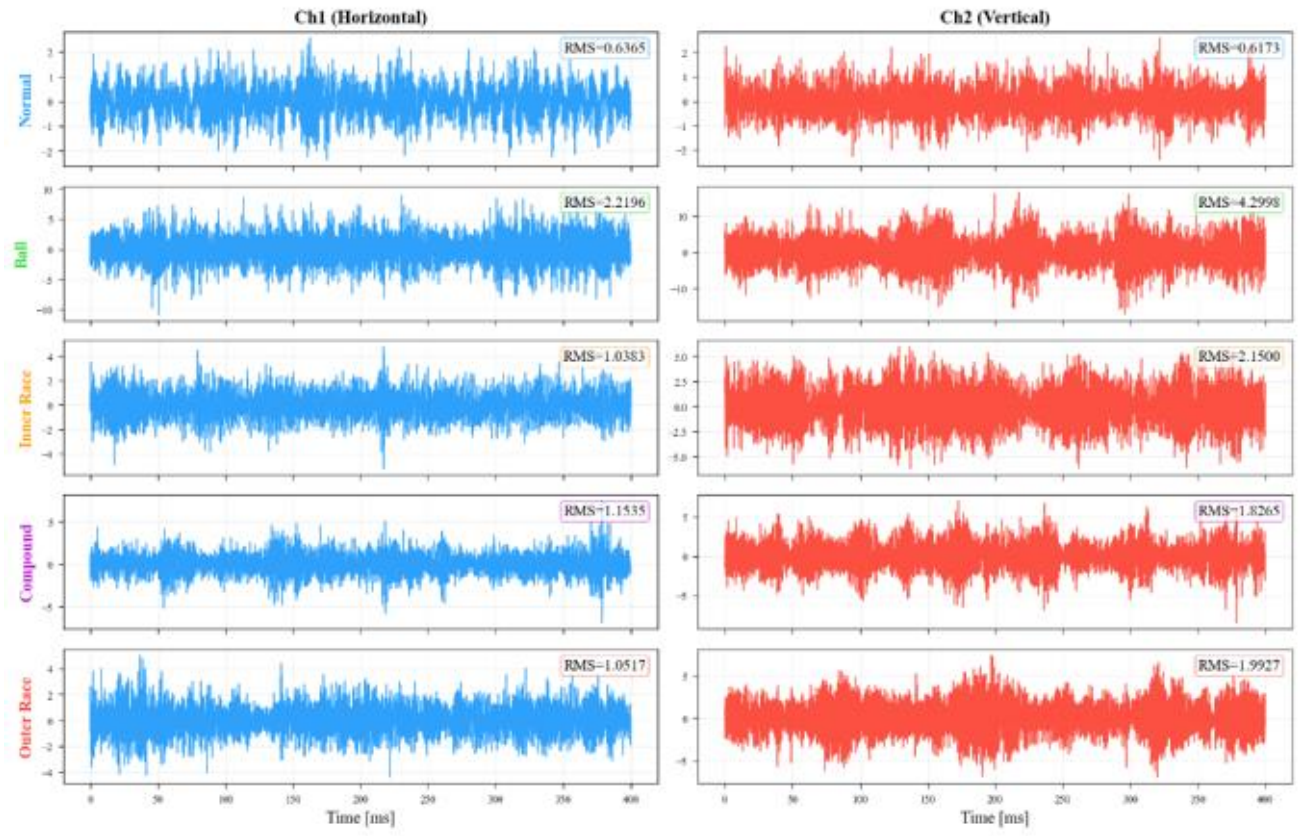

Fig. 2. Preprocessed vibration signals for both channels across all five fault classes. RMS values are annotated per segment.

## C. Characteristic Fault Frequencies

Theoretical defect frequencies are computed for the target bearing geometry at 1,800 r/min: Ball Pass Frequency Outer (BPFO) = 89.79 Hz, Ball Pass Frequency Inner (BPFI) = 120.21 Hz, Ball Spin Frequency (BSF) = 101.40 Hz, and Fundamental Train Frequency (FTF) = 12.83 Hz. These frequencies correspond to the periodicity of impact events when a rolling element contacts a defect on the outer race, inner race, ball surface, or cage, respectively. Knowledge of these frequencies motivates the multi-scale moment extraction strategy: different α exponents implicitly weight different regions of the amplitude distribution, providing sensitivity to both low-amplitude periodic events characteristic of Outer Race and Inner Race faults, and high-amplitude impulsive events characteristic of Ball and Compound faults).

# IV. PROPOSED METHODOLOGY

The PG-AMF pipeline consists of four stages, as illustrated in Fig 3: (i) adaptive moment feature extraction from each channel, (ii) diversity-regularized learning of exponent parameters, (iii) cross-channel bilinear fusion, and (iv) MLP classification.

## A. Adaptive Moment Feature Extraction

For a signal segment $x = \{x_1, \ldots, x^{\mathrm{T}}\}$ of length T, a generalized moment feature with learnable exponent $\alpha_{\mathrm{i}}$ is defined as:

$$f_i = ( 1/T )\ \Sigma t\ \ |x_t| \alpha_i \qquad (1)$$

Three moment families are computed per channel: (i) absolute moments using $|x_{\mathrm{t}}|$, capturing amplitude energy at multiple scales; (ii) signed moments using $|x_{\mathrm{t}}|^{\{\alpha_{\mathrm{i}}\}} \cdot sgn(x_{\mathrm{t}})$, preserving directional asymmetry relevant to Inner Race and Compound fault signatures; and (iii) AC-coupled moments replacing $x_{\mathrm{t}}$ with its mean-removed version, capturing oscillatory rather than DC-offset energy as observed in Outer Race periodic impact patterns. With 10 learnable exponents per family per channel, each channel produces a 30-dimensional feature vector; concatenation across two channels yields a 60-dimensional representation. Each $\alpha_i$ is initialized to a uniform spacing in [1, 2] and constrained to $\alpha_{\mathrm{i}} \geq 0.5$ via softplus parameterization.

## B. Diversity Regularization

Without regularization, learned exponents can collapse toward a single value, yielding redundant features. A margin-based diversity loss is imposed to ensure minimum separation between exponent values:

$$L_{div} = \Sigma i < j\ max(0, \delta - |\alpha_i - \alpha_j|) \qquad (2)$$

where $\delta = 0.1$ is the minimum separation margin. Pairs of exponents closer than $\delta$ are penalized, encouraging the framework to distribute learned moments across a range of sensitivity scales spanning from energy-based (low $\alpha$) to kurtosis-sensitive (high $\alpha$) descriptors.

## C. Cross-Channel Fusion

Let $h_1$ and $h_2$ denote the 30-dimensional feature vectors from $Ch_1$ and $Ch_2$, respectively. A bilinear interaction term $b = h_1 \odot W^{\mathrm{b}} h_2$ is computed, where $W^{\mathrm{b}} \in \mathbb{R}^{\{30\times 30\}}$ is a learnable weight matrix and $\odot$ denotes element-wise multiplication. A gated residual connection combines this term with the direct concatenation:

$$h_{fused} = \sigma(g) \odot [h_1;\ h_2] + (1 - \sigma(g)) \odot b \qquad (3)$$

where $g$ is a learnable scalar gate and $\sigma$ denotes the sigmoid function. The gate enables the model to balance direct concatenation, which preserves individual channel characteristics, with bilinear cross-channel interaction, which captures joint directional vibration patterns. This is particularly beneficial for distinguishing Compound faults from single-location faults, as compound defects produce simultaneous modulations across both horizontal and vertical axes.

## D. Classifier Architecture

The fused 60-dimensional feature vector is passed through batch normalization followed by a two-layer multilayer perceptron (MLP): 60 → 128 → 64 → 5 (number of fault classes), with ReLU activations and dropout (p = 0.3) between layers. The total model contains 44,065 trainable parameters, which is approximately 4.3 times that of the statistical baseline (10,245), but remains negligible compared to typical CNN architectures. The PG-AMF feature extraction introduces additional scalar parameters, representing minimal computational overhead relative to the classification head.

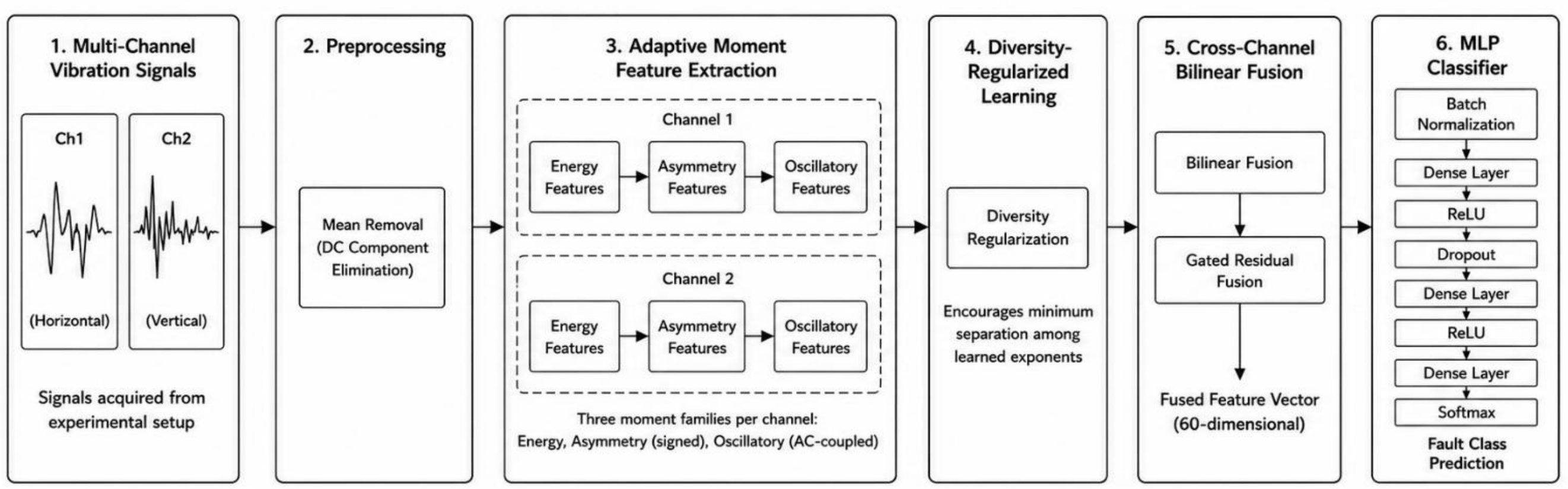

Fig. 3. PG-AMF framework for bearing fault diagnosis

## E. STATISTICAL BASELINE

A fixed-feature statistical baseline is implemented for comparison. Six hand-crafted features are extracted per channel: RMS, kurtosis, skewness, crest factor, variance, and peak-to-peak amplitude. The features from both channels are concatenated to form a 12-dimensional vector, which is fed into an MLP with architecture identical to the PG-AMF classifier (128 → 64 → 5). The baseline is trained using the Adam optimizer and standard cross-entropy loss. This baseline is designed to isolate the benefit of adaptive moment learning from any differences attributable to classifier architecture.

### F. *TRAINING STRATEGY*

A unified training strategy is adopted for both methods, with method-specific configurations detailed in Table I. Training is performed using mini-batch stochastic gradient descent with early stopping based on validation accuracy, and a fixed random seed is used to ensure reproducibility.

For the statistical baseline, a standard cross-entropy loss is minimized under a constant learning rate schedule. In contrast, the PG-AMF model is trained using a cosine annealing learning rate strategy and a composite loss function defined as:

$$L = L_{focal} + \lambda_{div}\, L_{div} + \lambda_{compact}\, L_{compact} \tag{4}$$

where the focal loss emphasizes hard-to-classify samples, the diversity regularization term promotes non-redundant feature representations, and the compactness loss improves intra-class clustering relative to inter-class separation. These additional objectives enable more discriminative and adaptive feature learning compared to the fixed statistical baseline. All experiments are conducted under identical training conditions to ensure a fair comparison, and the computational requirements remain modest, allowing efficient execution on standard CPU hardware.

TABLE I. TRAINING CONFIGURATION COMPARISON

| Hyperparameter | Stat. MLP (Baseline) | PG-AMF (Proposed) |
|---|---|---|
| **Optimizer** | AdamW | AdamW |
| **Initial LR** | 0.001 | 0.001 |
| **LR Schedule** | Constant | Cosine Annealing |
| **Loss Function** | Cross-Entropy | Composite (Focal + Diversity + Compactness) |
| **Early Stopping Patience** | 10 epochs | 10 epochs |
| **Max Epochs** | 200 | 200 |
| **Batch Size** | 64 | 64 |
| **Random Seed** | 42 | 42 |

## V. FEATURE DISCRIMINABILITY ANALYSIS

The multiclass Fisher Discriminant Ratio (FDR) is computed as:

$$FDR = t_r(S_W^{-1} S_B) \tag{5}$$

where $S_W$ is the within-class scatter matrix and $S_B$ is the between-class scatter matrix, both computed from the feature vectors. A higher FDR indicates greater inter-class separability relative to intra-class variance, which directly correlates with classification accuracy across Normal, Ball, Inner Race, Compound, and Outer Race classes. As illustrated in Fig. 4, the overall FDR for the statistical baseline across 12 features is 278.47, with $RMS_{Ch2}$ (134.20) and $Variance_{Ch2}$ (90.14) contributing the most discriminative power. In contrast, per-feature FDR values for kurtosis (Ch1: 0.91, Ch2: 1.07), crest factor (Ch1: 0.88, Ch2: 0.59), and skewness (Ch1: 2.11, Ch2: 0.80) are substantially lower than 5, indicating limited individual discriminability of these fixed-order descriptors for separating structurally similar fault classes such as Inner Race and Outer Race.

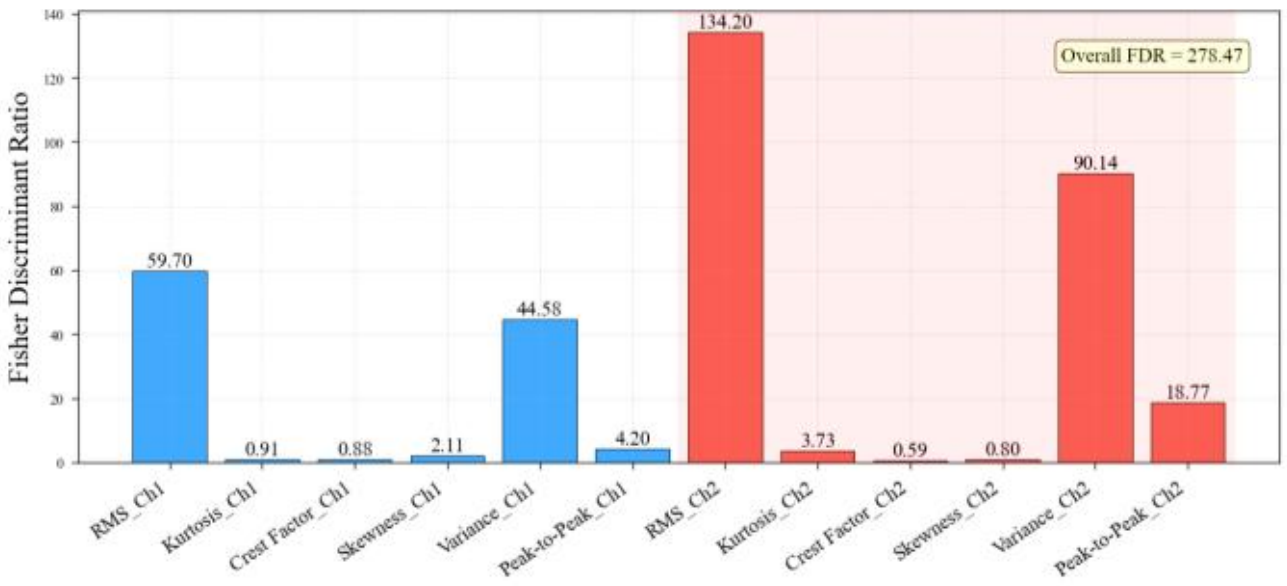

Fig. 4. Per-feature FDR for the statistical baseline.

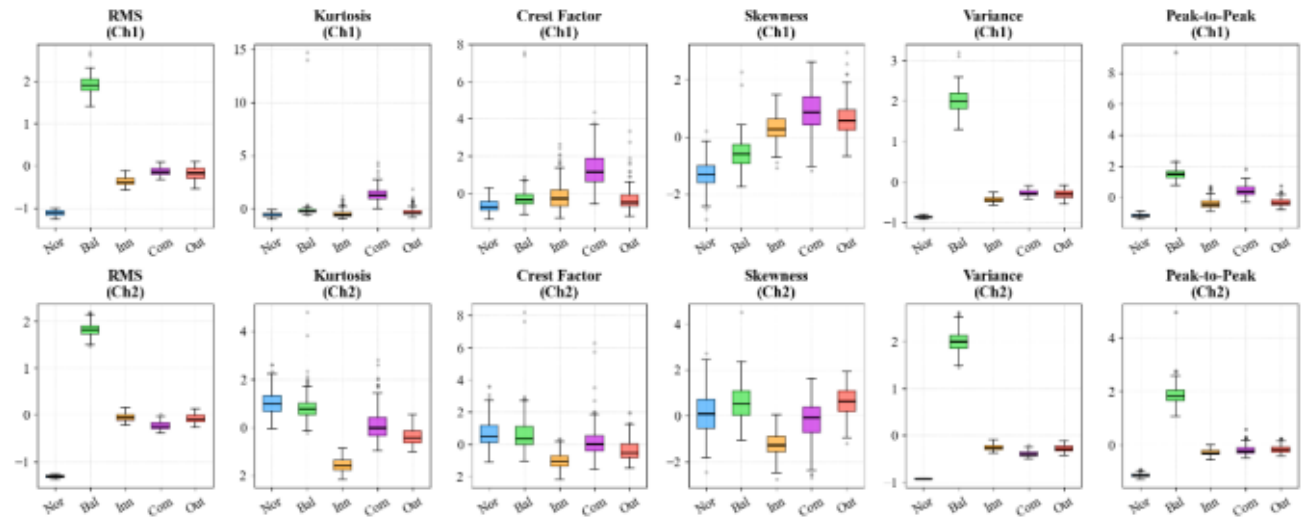

Fig. 5. Statistical feature distributions per fault class (top: Ch1; bottom: Ch2). Overlap between Inner Race, Compound, and Outer Race classes motivates adaptive exponent learning.

## V. RESULTS AND DISCUSSION

### A. *Learning Behavior*

The training and validation loss and accuracy curves for both methods are presented in Fig. 6. The statistical baseline converges rapidly, with the best validation accuracy of 100.0% reached at approximately epoch 22 during training, yet generalizes to only 96.0% on the held-out test set, indicating modest overfitting to the training distribution. The PG-AMF model, trained with cosine annealing and the composite loss, converges more gradually, with the optimal validation accuracy attained at epoch 54. Mild oscillation in the PG-AMF validation loss is attributable to the diversity regularization term, which actively perturbs the exponent landscape during optimization; however, the moving-average trend confirms stable convergence toward a discriminative solution.

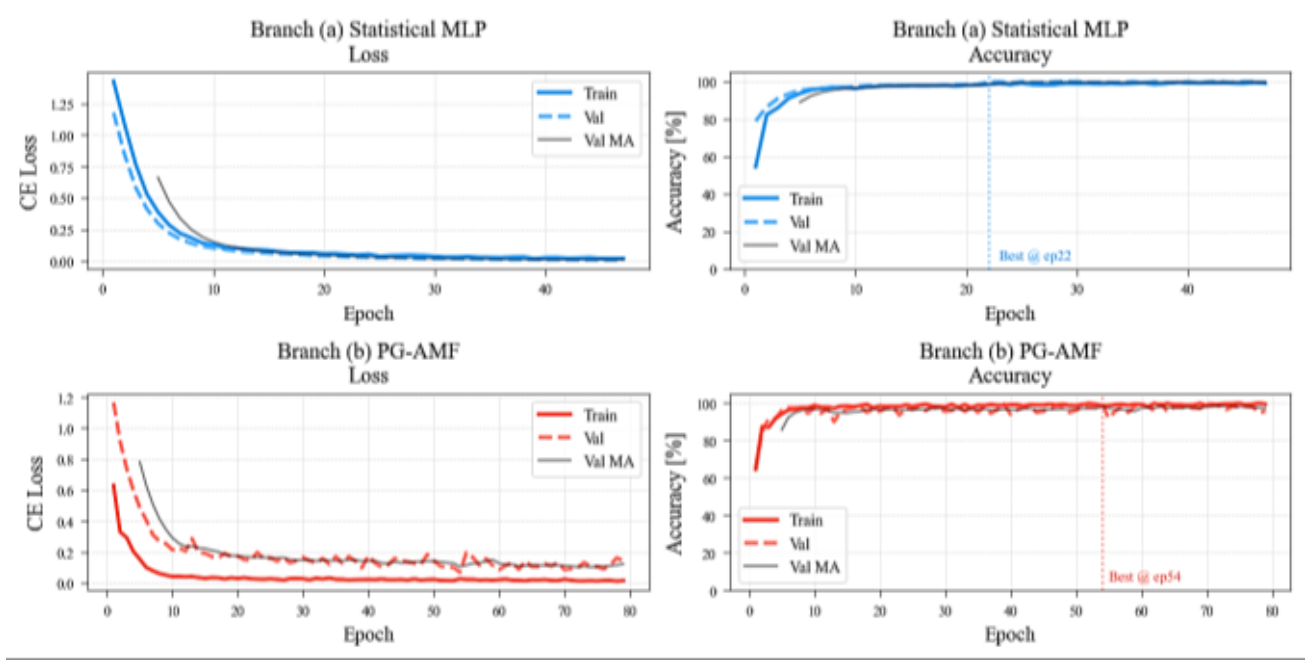

Fig. 6. Training and validation curves. Top: Statistical MLP. Bottom: PG-AMF with cosine annealing, focal loss, and diversity regularization.

## B. Classification Performance

As summarized in Table II, PG-AMF achieves perfect classification across all five fault classes, with an accuracy and Macro-F1 of 100.0% on the held-out test set. The statistical baseline achieves 96.0% accuracy. As shown in the normalized confusion matrices of Fig. 7, misclassifications in the baseline are confined to two classes: 3% of Inner Race fault samples are incorrectly assigned to the Outer Race class, and 12% of Compound fault samples are misclassified as Outer Race faults. The Compound class, which exhibits concurrent inner, outer, and ball fault signatures, poses the greatest challenge for fixed-feature methods due to the overlapping statistical distributions visible in Fig. 5. Normal, Ball, and Outer Race fault classes are correctly by both methods, confirming that these three classes are well-separated in the fixed-feature space; however, the subtle spectral modulations that distinguish Inner Race, Compound, and Outer Race faults require the adaptive exponent learning afforded by PG-AMF.

TABLE II. TEST PERFORMANCE FOR BASELINE AND PROPOSED METHOD

| Method | Accuracy | Macro-F1 |
|---|---|---|
| **Statistical MLP (baseline)** | 96.00% | 96.01% |
| **PG-AMF (proposed)** | 100.00% | 100.00% |

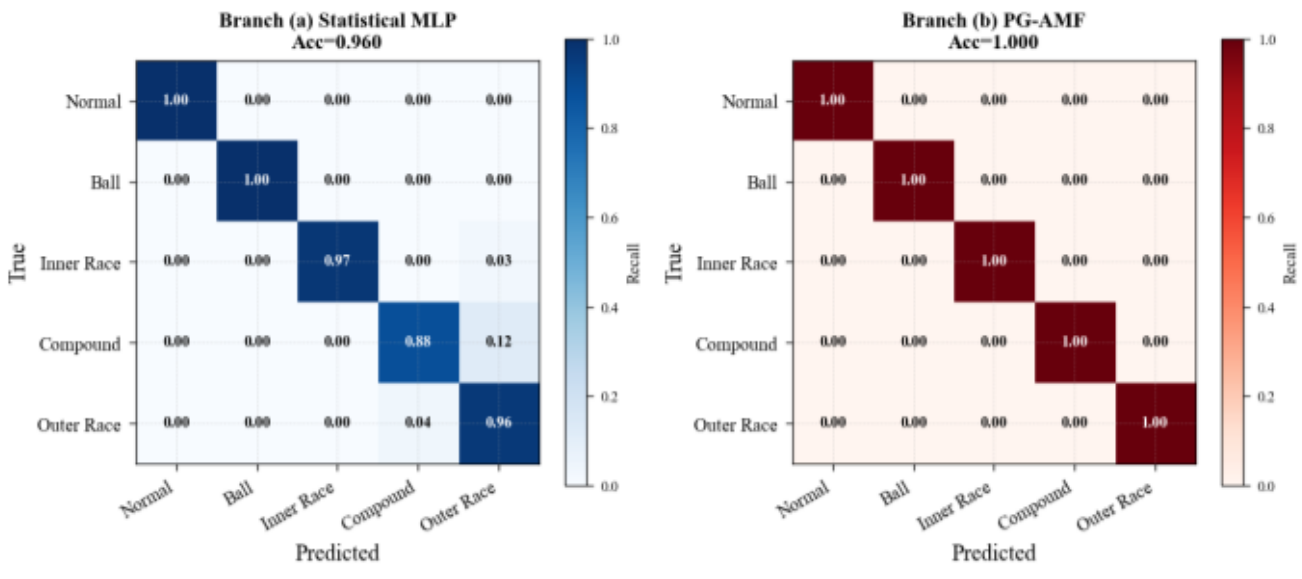

Fig. 7. Normalized confusion matrices on the held-out test set. Left: Statistical MLP. Right: PG-AMF. Misclassifications are confined to Inner Race and Compound in the baseline.

## C. Learned Exponent Analysis

The distribution of learned α values across all 10 moment indices and both channels, as illustrated in Fig. 8, reveals that exponents span the range [1.05, 4.47], forming a near-linear progression that covers classical moment orders. Reference lines at $\alpha = 2$ and $\alpha \approx 4$ correspond to RMS energy (second-order moment) and kurtosis-analogue sensitivity (fourth-order moment), respectively. The convergence of both Ch_1 and Ch_2 to nearly identical α profiles confirms that the diversity regularization successfully prevents channel-specific collapse, and that the multi-family extraction captures qualitatively distinct aspects of the vibration signal. The low-α exponents (indices 0–2) are sensitive to broadband energy and are most informative for separating Normal from all fault classes; the high-α exponents (indices 7–9) capture impulsive characteristics relevant to Ball and Inner Race fault discrimination. This interpretable structure establishes a direct physical link between learned features and established vibration analysis theory.

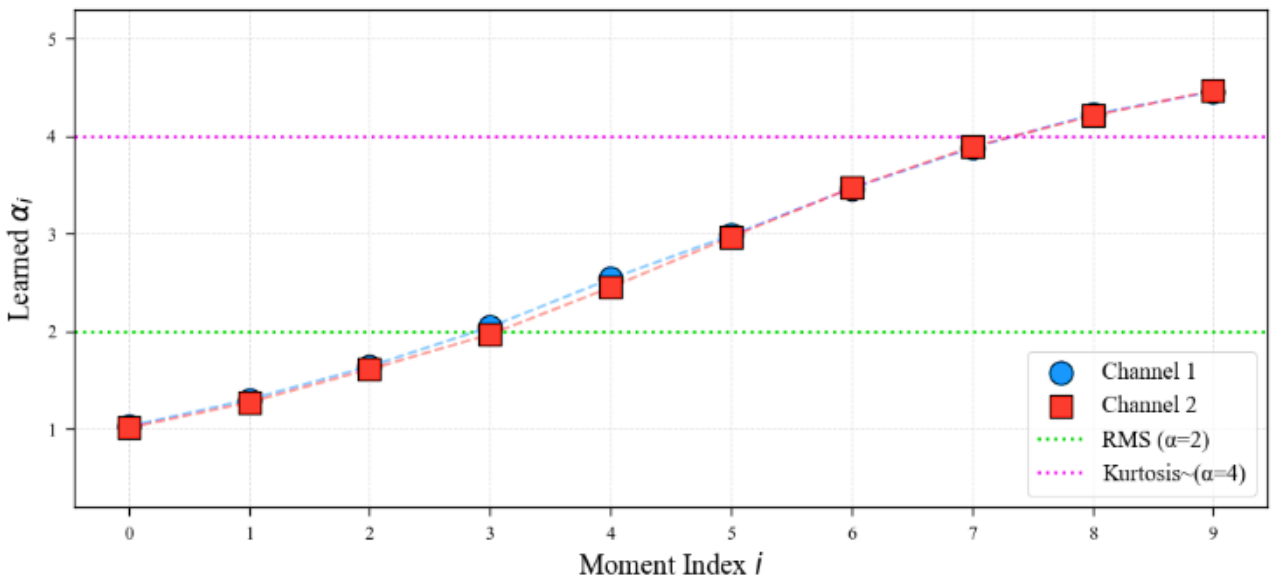

Fig. 8. Learned PG-AMF exponents α_i after training. Reference lines at α=2 (RMS) and α≈4 (kurtosis analogue) are shown

## D. Post-Training t-SNE Visualization

The t-SNE projections of the feature space after training are presented in Fig. 9. For the statistical baseline, well-separated clusters are observed for the Normal and Ball fault classes; however, the Inner Race, Compound, and Outer Race feature clusters exhibit partial spatial overlap, which is consistent with the misclassification pattern observed in Fig. 6. In the PG-AMF feature space, all five fault classes form compact and well-separated clusters. The Normal class occupies a highly elongated linear manifold, reflecting the regularity of the healthy vibration signal under the learned adaptive exponents. The Ball fault class is projected to a region spatially distant from all other classes, consistent with its characteristically large RMS amplitudes across both channels. The Inner Race and Outer Race classes, which are most frequently confused in the baseline, are projected to non-overlapping regions in the PG-AMF space, confirming the effectiveness of adaptive exponent learning for fine-grained fault discrimination.

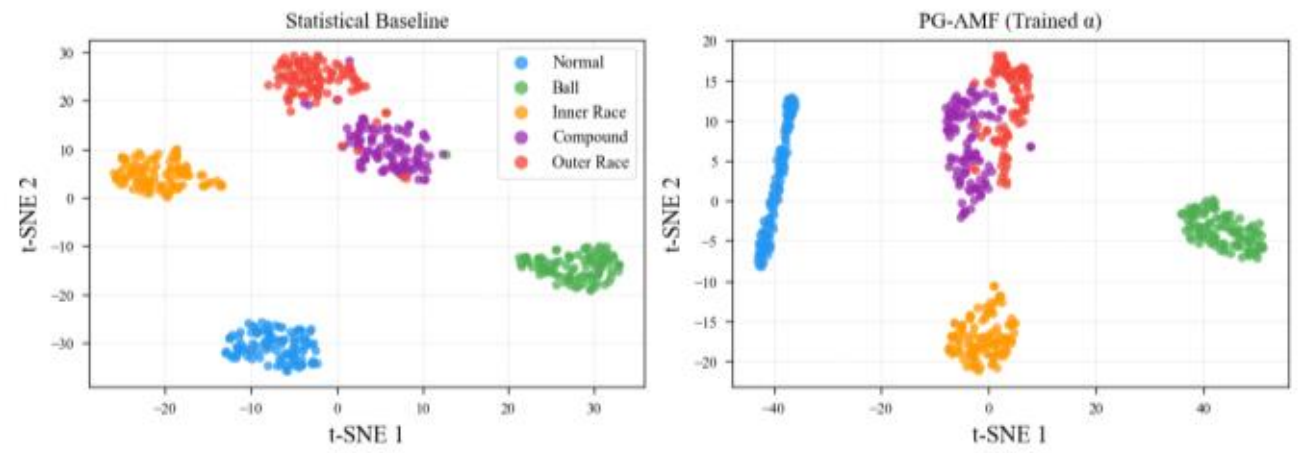

Fig. 9. Post-training t-SNE projections. Left: Statistical baseline. Right: PG-AMF. PG-AMF features exhibit tighter intra-class clusters and wider inter-class margins.

## E. Cross-validation

Five-fold stratified cross-validation is conducted to assess statistical stability. As reported in Table III, PG-AMF achieves a mean accuracy of 99.53 ± 0.69% and mean Macro-F1 of 99.53 ± 0.68% across five folds, compared to 96.82 ± 0.83% and 96.82 ± 0.84% for the statistical baseline. The reduced standard deviation indicates that PG-AMF is not only more accurate but also more consistent across data partitions. As detailed in Table IV, the Normal class achieves 100.0 ± 0.0% for PG-AMF compared to 98.0 ± 0.0% for the baseline. Ball fault F1 is significantly improved from 96.7 ± 0.6% to 99.7 ± 0.6%, reflecting the distinctive amplitude signature of ball defects. The most notable gains are observed for Inner Race faults, where PG-AMF achieves 100.0 ± 0.0% F1 compared to 95.4 ± 1.2% for the baseline, and for Outer Race faults (99.1 ± 1.2% versus 95.4 ± 2.2%), where the higher baseline variance reflects the structural similarity between Outer Race and Compound fault signatures in the fixed-feature space.

TABLE III. 5-FOLD CROSS-VALIDATION (MEAN ± STD)

| Method | Accuracy | Macro-F1 |
|---|---|---|
| **Statistical MLP** | 96.82% ± 0.83% | 96.82% ± 0.84% |
| **PG-AMF** | 99.53% ± 0.69% | 99.53% ± 0.68% |

TABLE IV. PER-CLASS F1 (MEAN ± STD WITH 5-FOLD CV)

| Method | Normal | Ball | Inner Race | Compound | Outer Race |
|---|---|---|---|---|---|
| **Statistical MLP** | 98.0±0.0% | 96.7±0.6% | 95.4±1.2% | 95.6±1.5% | 95.4±2.2% |
| **PG-AMF** | 100.0±0.0% | 99.7±0.6% | 100.0±0.0% | 98.8±1.7% | 99.1±1.2% |

The principal mechanism by which PG-AMF achieves improved classification is the gradient-driven optimization of moment exponents, which adapts feature sensitivity to the specific amplitude distribution of each fault class. Fixed-order descriptors implicitly commit to a single region of the sensitivity spectrum; in contrast, the diversity-regularized exponent set spans the range from energy-sensitive (low α, relevant to Normal and Ball conditions) to impulsive-sensitive (high $\alpha$, relevant to Inner Race and Outer Race localized defects). The bilinear cross-channel fusion captures joint modulations of horizontal and vertical vibration components, which are particularly informative for Compound faults that produce simultaneous signatures at the BPFI, BPFO, and BSF frequencies. The per-class improvements in Table IV and Fig. 7 are consistent with this mechanism: Inner Race and Compound faults, whose fixed-feature representations overlap most strongly as evidenced in Fig. 5, benefit most from adaptive exponent learning, with PG-AMF achieving perfect Inner Race classification in all five folds.

An accuracy improvement from 96.0% to 100.0% is observed on the held-out test set, corresponding to the elimination of all misclassifications in the Inner Race and Compound fault classes. The cross-validation standard deviation is reduced from 0.83% to 0.69%, confirming that the gain is consistent and not attributable to favourable data partitioning. The FDR is elevated from 278.47 in the fixed-feature space to over 1,500 in the PG-AMF feature space post-training, quantifying the substantially improved inter-class separability across all five bearing health states. Despite containing approximately 4.3 times as many parameters as the baseline, PG-AMF remains a compact model with 44,065 total parameters and negligible inference cost relative to CNN-based alternatives.

The learnable exponent structure provides an interpretable record of the statistical moments upon which the diagnostic decision is based. The learned α distribution can be inspected by maintenance engineers to verify that the model relies on physically meaningful signal statistics, such as the RMS-analogue and kurtosis-analogue moment orders identified in Fig. 8. This interpretability is of value in safety-critical industrial applications where model certification and explainability are regulatory requirements. The parameter efficiency and CPU-only inference, requiring less than 500 milliseconds per fold on standard hardware, indicate suitability for deployment in edge-computing predictive maintenance platforms without specialized accelerator hardware.

Several limitations of the present work are acknowledged. First, all experiments are conducted under a constant rotational speed of 1,800 r/min; generalization to variable-speed operating conditions, in which the BPFO, BPFI, BSF, and FTF frequencies shift continuously, has not been validated and constitutes a priority for future investigation. Second, the experimental dataset comprises 1,000 balanced segments; the behavior of PG-AMF under severe class imbalance or highly limited labeled data requires further study. Third, although the scalar exponents α are interpretable in terms of moment order, the bilinear fusion weight matrix $W_b$ does not admit a direct physical interpretation in terms of sensor placement geometry or directional fault propagation. Fourth, the training time for PG-AMF, ranging from 143 to 458 seconds per fold on CPU, is substantially greater than that of the statistical baseline, which converges within 1.3 to 2.0 seconds per fold, owing to the richer composite loss computation and the gradient propagation through the learnable exponents.

## VI. CONCLUSION

The PG-AMF framework for vibration-based bearing fault diagnosis has been presented, in which moment exponents are treated as learnable parameters co-optimized with a composite loss function. On the XJTU Gearbox bearing dataset, PG-AMF achieves an accuracy and Macro-F1 of 100.0% on the held-out test set, and 99.53 ± 0.69% accuracy over five-fold cross-validation, compared to 96.0% and 96.82 ± 0.83% for the fixed statistical baseline. All five fault classes, including Normal, Ball, Inner Race, Compound, and Outer Race conditions, are correctly classified on the test set. The learned exponent distribution spans $\alpha \in [1.05, 4.47]$, covering classical RMS and kurtosis-analogue descriptors and providing an interpretable link to established vibration analysis theory. The overall FDR is substantially elevated post-training, and t-SNE projections confirm tighter cluster compactness for all fault classes. The framework is parameter-efficient, CPU-deployable, and compatible with standard two-channel accelerometer setups. Extension of PG-AMF to variable-speed operating conditions, evaluation of cross-machine type transfer, and integration with neuromorphic or spiking neural network architectures for ultra-low-power embedded deployment are planned for future work.